\documentclass{aa}
\usepackage[dvips]{graphicx}
\def\simless{\mathbin{\lower 3pt\hbox
{$\rlap{\raise 5pt\hbox{$\char'074$}}\mathchar"7218$}}}   %< or of order
\def\simmore{\mathbin{\lower 3pt\hbox
{$\rlap{\raise 5pt\hbox{$\char'076$}}\mathchar"7218$}}}   %> or of order
\newcommand{\be}{\begin{equation}}
\newcommand{\ee}{\end{equation}}

\begin{document}
\title{The role of kink instability in Poynting-flux dominated jets}
\titlerunning{The role of kink instability in jets}
\author{Dimitrios Giannios \and Henk C. Spruit }

\institute{Max Planck Institute for Astrophysics, Box 1317, D-85741 Garching, Germany}

\offprints{giannios@mpa-garching.mpg.de}
\date{Received / Accepted}

\abstract
{The role of kink instability in magnetically driven jets is explored
through numerical one-dimensional steady relativistic MHD calculations. The instability
is shown to have enough time to grow and influence the dynamics of Poynting-flux dominated
 jets. In the case of AGN jets, the flow becomes kinetic flux dominated at distances 
$\simmore 1000 r_{\rm{g}}$ because of the rapid dissipation of Poynting flux. When applied 
to GRB outflows, the model predicts more gradual Poynting dissipation and moderately magnetized 
flow at distances of $\sim10^{16}$ cm where the deceleration of the ejecta due to interaction
with the external medium is expected. The energy released by the instability can power the compact
``blazar zone'' emission and the prompt emission of GRB outflows with high radiative
efficiencies. \keywords{Magnetohydrodynamics (MHD) -- Instabilities -- Gamma rays: bursts -- Quasars: general }}

\maketitle

\section{Introduction} 
\label{intro}

Relativistic collimated outflows have been extensively observed in active galactic nuclei
(AGN) and X-ray binaries (XRB). Gamma-ray 
bursts (GRB) are also believed to be connected to ultrarelativistic and collimated outflows
to overcome the ``compactness problem'' (e.g. Piran 1999) and to explain the
achromatic afterglow breaks (Rhoads 1997; Sari et al. 1999). It is also believed that
all these sources are powered by accretion of matter by a compact object. 

The widely accepted mechanism for jet acceleration and collimation in the context of AGN and
XRB jets is that of magnetic driving. According to this paradigm, magnetic fields 
anchored to a rotating object can launch an outflow.
The rotating object can be a star (Weber \& Davis 1967; Mestel 1968), a pulsar (Michel 1969; 
Goldreich \& Julian 1970),
an accretion disk (Bisnovatyi-Kogan \& Ruzmaikin 1976; Blandford 1976; Lovelace 1976; 
Blandford \& Payne 1982) or a rotating black hole (Blandford \& Znajek 1977).
The material is accelerated thermally up to the sonic point and centrifugally until the 
Alfv\'en point, defined as the point where the flow speed equals the Alfv\'en speed. After the
Alfv\'en point the inertia of mater does not allow corotation of the magnetic field. As a
result, the magnetic field lines bend, developing a strong toroidal component.

Further out the flow passes through the fast magnetosonic point where most of the energy
of the flow remains in the form of Poynting flux in the case of relativistic outflows (Michel 1969;
Goldreich \& Julian 1970; Sakurai 1985; Beskin et al. 1998). 
Further acceleration of the flow is not straightforward within ideal MHD. It can be shown, for
example, that a radial flow is not accelerated after the fast point (e.g. Beskin 1998).
This is a result of the fact that the magnetic pressure and tension terms of the Lorentz force 
almost cancel each other (Begelman \& Li 1994).   A limited degree of acceleration
of the flow  is possible if it has a decollimating shape (i.e. the magnetic field 
diverges faster than radial; Li et al. 1992; Begelman \& Li 1994).

Magnetized jets suffer from a number of instabilities.  Interaction with the environment 
causes instabililty of the Kelvin-Helmholtz type and kink instability causes internal rearrangement
of the field configuration.
Here we focus   on kink instability, since it internally dissipates magnetic energy 
associated with the Poynting flux. As demonstrated elsewhere
(Drenkhahn 2002; Drenkhahn \& Spruit 2002; Spruit \& Drenkhahn 2003) such
internal energy dissipation directly leads to acceleration of the flow. Dissipation  
steepens the radial decrease of magnetic pressure, thereby lifting the cancellation
between outward pressure force and inward magnetic tension, and allowing the magnetic
pressure gradient to accelerate the flow.

\subsection{``AC'' versus ``DC'' outflows} 

While dissipation of magnetic energy can thus happen through kink instability in an 
initially axisymmetric (``DC'') flow, it can also happen more directly by reconnection in the 
outflow generated by a non-axisymmetric rotator (``AC'' flow). The two cases behave 
differently in terms of the acceleration profile, and the location and amount of 
radiation produced by the dissipation process.  A nonaxisymmetric rotator produces
a ``striped'' outflow (as in the case of a pulsar wind) with reconnectable changes of
direction of the field embedded in the flow, and energy release  independent of
the opening angle of the jet. In the DC case, where energy release is instead 
mediated by an instability, the rate of energy release is limited by the time it takes an
Alfv\'en wave to travel across the width of the jet. This makes it a sensitive function
of the jet opening angle. 

The ``AC'' case has been studied in detail by Drenkhahn (2002), Drenkhahn and Spruit 
(2002), with application to Gamma-ray bursts. In the case of AGN and XRB, on the other 
hand, the collimated jet is arguably best understood if the field in the inner 
disk is of uniform polarity, resulting in an initially axisymmetric flow. Another difference
is the lower bulk Lorentz factors in the AGN/XRB case, resulting in faster energy release
(in units of the dynamical time of the central engine).

The purpose of this paper is to explore the consequences of  magnetic dissipation by 
internal instability in such axisymmetric (or DC) cases, and its observational signatures.
We also apply the calculations to the GRB case, where we compare the results with the AC 
case studied  before. 

\subsection{Energy release and field decay by the instability}

We limit ourselves to a flow with constant opening angle. That is, we leave aside the
collimation process. Kink instability is modeled by adding a sink term to the induction 
equation to account for the non-ideal MHD effects arising from it.

  Linear stability 
theory of kink instability yields a growth time of the order $t_{\rm k}=r\theta /v_{\rm A,\phi}$ 
where $r$ is the radius of the jet, $\theta$ its opening angle and $v_{\rm A,\phi}$ 
the Alfv\'en speed based on the 
azimuthal component $B_\phi$ of the field. This is independent of the poloidal field 
component, at least for the so-called internal kink modes (which do not disturb the outer 
boundary of the field, cf.\  Bateman 1978 for details). For stability analysis and numerical 
simulations with astrophysical applications see, e.g. Begelman (1998); Appl et al. (2000); 
Lery et al. (2000); Ouyed et al. (2003); Nakamura \& Meier (2004). Based on linear
theory, we would predict that the poloidal field component
can be ignored for the rate of energy release. 

It is not clear, however, that the poloidal
component can be ignored for the nonlinear development of the instability, which is
what actually determines the energy release. As a way to explore the effect
of possible nonlinear stabilization by a poloidal component we compare two cases
in the calculations: one with energy release and field decay given by the Alfv\'en time
across the jet ( $t_{\rm k}$ above), and one in which this rate is assumed to be reduced 
by the poloidal component. This mainly affects the early phases of the acceleration of the 
flow beyond the light cylinder. 
   
 We find that kink instability has time to grow  in the AGN and
XRB cases,  dissipating energy in the toroidal component of the magnetic field  while
accelerating the flow at the same time. The dissipation of magnetic energy 
is almost complete and fast in the case of AGN jets,  so that on parsec scales
the flow has become kinetic energy dominated, in agreement with current interpretations
of the observations (e.g. Sikora et al. 2005, where the possible effects of magnetic dissipation
are also discussed briefly).  

The DC model with kink instability also produces significant flow acceleration
in the GRB case, but conversion of the Poynting flux is less effective than the AC model in
this case. 

The structure of the paper is as follows. In Sec.~2 we discuss MHD instabilities
in jets and focus on the kink instability and its growth rate. The model is described in Sec.~3
including the assumptions, the dynamical equations and the parameters at the base of the flow.
In Sec.~4, we apply the model to the case of AGN jets and GRBs, while the last two Sections
present the discussion and conclusions.

\section{The kink instability}
\label{kink}

Magnetized outflows are subject to a variety of instabilities. These can be classified as 
pressure driven, Kelvin-Helmholtz and current driven instabilities (see, e.g., Kadomtsev 1966;
Bateman 1978). Pressure driven 
instabilities (Kersal\'e et al. 2000; Longaretti 2003) are related to the interplay between the 
gas pressure and the curvature of magnetic field lines. They are relevant close to the launching region 
of the outflows and may be important as long as the outflow is still subsonic. Kelvin-Helmholtz (KH) 
instabilities (Ray 1981; Ferrari 
et al. 1981; Bodo et al. 1989; Hardee \& Rosen 1999) arise from 
velocity gradients in the flow and may be important in the shearing layer between the outflow and the 
external medium.  KH instabilities have been extensively studied and  become strongest
in the region beyond the Alfv\'en point but still within the fast magnetosonic point. Current driven 
(CD; Eichler 1993; Spruit et al. 1997; Begelman 1998; Lyubarskii 1999; Appl et al. 2000) 
instabilities have received much less attention but  are the most relevant ones for 
Poynting-flux dominated outflows,  since they can convert the bulk Poynting flux into radiation and
kinetic energy of the flow (for the role of CD instabilities in an electromagnetic model
for GRBs see Lyutikov \& Blandford 2003). Among the CD instabilities, the $m=$1 kink instability 
is generally the most effective. In this work, we focus on the effect of the kink instability 
on the dynamics of these outflows. 

\subsection{The growth rate of the instability}
 
While magnetized outflows can be  accelerated ``centrifugally''  by large scale poloidal fields 
(Blandford \& Payne 1982; Sakurai 1985, 1987), at the radius of the light-cylinder inertial forces 
become significant and the magnetic field 
cannot force corotation. At this radius the strength of the toroidal and the poloidal 
components are comparable. Further out, the induction equation dictates that, within ideal MHD, 
the toroidal component dominates over the poloidal one since the strength of the former
scales as $1/r$ while that of the latter as $1/r^2$. This magnetic configuration of a strongly 
 wound-up magnetic field  like this is known, however, to be highly unstable to the kink $m=1$ mode
from tokamak experiments (see, e.g., Bateman 1978). Linear stability analysis 
has shown that the growth time of the instability is given by the Alfv\'en crossing time across the 
outflow in a frame comoving with it (Begelman 1998; Appl et al. 2000).
  
The study of the non-linear evolution of the instability demands three dimensional relativistic
MHD simulations over many decades of radii and it is, therefore, not surprising that the issue is not
settled. Lery et al. (2000) and Baty \& Keppens (2002) argued in favor of the dynamical importance 
of the instability in reorganizing the magnetic configuration inside the jet. It has been argued, 
however,  that the jet creates a ``backbone'' of strong poloidal field which slows 
down the development of instabilities (Ouyed et al. 2003; Nakamura \& Meier 2004). In view of these 
works and since the growth rate of the instability is important for this study, we   
consider two alternatives for the non-linear stage of the instability. In the first case, the 
instability proceeds at the Alfv\'en crossing time across the outflow (as suggested by linear stability
 analysis) and rearranges the magnetic field configuration to a more chaotic one. In this case the 
instability time scale is given by the expression (in the central engine frame)
\be
t_{\rm{k}}=\frac{r\theta \gamma}{v_{\rm{A,\phi}}}.
\label{kink1} 
\ee
We will refer to this as the ``fast kink'' case. 

 For the second case, we reduce the dissipation rate by a suitable (but arbitrary)
function of the poloidal-to-toroidal field ratio. 
\be     
t_{\rm{k}}=\frac{r\theta \gamma}{v_{\rm{A,\phi}}}e^{B^{\rm{co}}_p/B^{\rm{co}}_\phi}.
\ee
We will refer to this as the ``slow kink'' case.  This recipe is meant only as a means 
to explore the possible effect that the poloidal field component could have on the net acceleration
of the flow, if it affects the dissipation rate, and is not meant to be quantitative.
Numerical simulations of the instability
{would be} needed to determine which of these prescriptions (if any) of the growth time scale is 
close in describing its non-linear development (see also Section 5).
In the last expressions, $B^{\rm{co}}_\phi$,  $B^{\rm{co}}_p$, are the toroidal and the poloidal 
components of the magnetic field as measured by an observer comoving with the flow, $\theta$ is 
the jet opening angle, $\gamma$ is the bulk Lorentz factor of the flow and $v_{\rm{A,\phi}}$ 
is the $\phi$ component of the Alfv\'en speed given by
\be
v_{\rm{A,}\phi}=c\frac{u_{\rm{A},\phi}}{\sqrt{1+u_{\rm{A,}\phi}^2+u_{\rm{A,}p}^2}}, \qquad 
u_{\rm{A,}\phi}=\frac{B_\phi^{\rm{co}}}{(4\pi w)^{1/2}}.
\label{alfvenspeed}
\ee 
Here, $w$ is the internal enthalpy, to be defined below. 

\section{The model}

A magnetically launched outflow passes through three characteristic points where the speed of the flow 
equals the speed of slow mode, the poloidal Alfv\'en wave and the fast mode and are called the 
slow magnetosonic, the Alfv\'en and the fast magnetosonic points respectively. For flows where the 
energy density of the magnetic field dominates that of matter, the Alfv\'en point lies very close
to the light-cylinder 
\be
R_L=c/\Omega,
\ee
where $\Omega$ is the angular velocity of the foot point (e.g. Camenzind 1986).
At the Alfv\'en radius most of the centrifugal acceleration has already taken place and
the magnetic field cannot force corotation of matter. At this location, the toroidal and the poloidal 
components of the magnetic field are comparable in magnitude.  
Further out, the flow passes through the fast magnetosonic point at a distance
$\sim \rm{a \quad few}\quad R_L$ (Sakurai 1985; Li et al. 1992; Beskin et al. 1998). 
At the location of the fast 
magnetosonic point the speed of the four-velocity of the flow equals $\sim \mu^{1/3}$,
where $\mu$ is the Michel magnetization parameter (i.e., the energy flux per unit rest mass; Michel 1969). 
For Poynting-flux dominated flows (i.e., $\mu\gg 1$), most of the energy is still in magnetic form at this
point since the ratio of magnetic to matter energy flux is $\sim \mu^{2/3}$. 
 There is thus a choice between flows with high Lorentz factors (but inefficient
conversion of Poynting flux to kinetic energy), or efficient conversion at
the price of low terminal Lorentz factors. Better conversion within ideal MHD
appears to be hard to achieve except by decollimation of the flow (Li et al. 1992;
Begelman and Li 1994, Bogovalov 2001; Daigne and Drenkhahn 2002; but see claims to the 
contrary by Vlahakis and Konigl 2003a,b; Fendt and Ouyed 2004). Even with such
decollimation, the additional acceleration is rather modest (Begelman and Li 1994; Daigne and
Drenkhahn 2002).

We set the initial conditions of our calculation at the fast magnetosonic point $r_0$. To make 
the problem tractable we make a number of simplifying assumptions. First, we limit ourselves to 
a {\it radial}, {\it static} flow. Evidently, this approach does not allow us to explore the important issue 
of jet collimation (see, however Section 5.1). Furthermore, the flow is assumed {\it one-dimensional}
by ignoring the structure of the jet in the $\theta$ direction. Also, we ignore 
the azimuthal component of the velocity. This 
component is not dynamically important beyond the fast magnetosonic point (e.g. Goldreich \& 
Julian 1970) and can be neglected from the dynamic equations. On the other hand, the poloidal 
component (taken to be radial for simplicity) still has to be taken into account when modeling 
the effect of the kink instability since it influences its growth timescale [see Eqs.~(1), (2)].
These simplifying assumptions minimize the number of the free parameters of the model, 
allowing us to study the effect of each on the jet dynamics, as will become clear 
in the next sections.
 
\subsection{Dynamical equations}

To determine the characteristics of the flow as a function of radius, one needs the conservation
equations for mass, energy and angular momentum. These equations can be brought in the form
(if, for the moment, we neglect radiative losses; e.g. Lyutikov 2001; Drenkhahn 2002)
\be
\partial_rr^2\rho u=0,
\label{massc}
\ee  \be
\partial_rr^2\Big(w\gamma u+\frac{\beta B_\phi^2}{4\pi}\Big)=0,
\label{energyc}
\ee \be
\partial_r r^2\Big(wu^2+p+\frac{(1+\beta^2)B_\phi^2}{8\pi}\Big)=2rp,
\label{momentumc}
\ee 
where $w=\rho c^2+e+p$ is the proper enthalpy density, $e$ and $p$ are the internal energy 
and pressure respectively and $u=\gamma \beta$ is the radial four-velocity. We still need to 
assume an equation of state that will provide a relation between the pressure $p$ 
and the internal energy. Assuming an ideal gas, we take $p=(\gamma_a-1)e$,
where $\gamma_a$ is the adiabatic index. 

Mass conservation (\ref{massc}) can be integrated to yield mass flux per sterad 
\be 
\dot{M}=r^2u\rho c, 
\ee
while energy conservation gives the total luminosity per sterad
\be
L=wr^2\gamma u c+\frac{\beta c(rB_\phi)^2}{4\pi}.
\label{lum}
\ee
The first term of the last expression corresponds to the kinetic energy flux and the second
to the Poynting flux. 
A key quantity is the ``magnetic content'' of the flow which we will refer
to as the magnetization parameter $\sigma$, defined as the ratio of the radial Poynting to matter energy flux
\be
\sigma=\frac{L_{\rm pf}}{L_{\rm{kin}}}=\frac{B_\phi^2}{4\pi \gamma^2w}.
\label{sigma}
\ee  

For a flow to reach large asymptotic Lorentz factors (observations indicate $\gamma \sim 10-20$ 
for quasars, and theoretical arguments arising from the ``compactness problem'' such as Piran 1999 
constrain $\gamma \simmore 100$ for GRBs), it must start with a high energy to mass ratio. Within 
the fireball model for GRBs (Paczy\'nski 1986; Goodman 1986)
this means that $e\gg \rho c^2$. In this work, we focus on the opposite limit where most of the energy
 is initially stored in the magnetic field ($\sigma_0\gg 1$), while we treat the flow as cold (i.e., 
$e\simless \rho c^2$). Obviously, there can exist an intermediate regime of a
``magnetized-fireball'' models where both $e/\rho c^2$ and $\sigma_0\gg 1$ at the base of the flow.

Finally, the strength of the radial component is given by flux conservation by the expression
\be
B_r=B_{r,0}\Big(\frac{r_0}{r}\Big)^2.
\label{br}
\ee  
For a flow that is moving radially with a bulk Lorentz factor $\gamma$, the expressions that relate
the comoving components of the magnetic field to those measured in the central engine frame
are
\be
B^{co}_r=B_{r}
\label{brlabco}
\ee
and
\be
B^{co}_\phi=\frac{B_\phi}{\gamma}.
\label{bphilabco}
\ee         
 
\subsection{Modeling the kink instability}

The set of equations presented in the previous section is not complete. There is one more 
equation needed to determine the problem at hand, which is the induction equation. For ideal
MHD, the induction equation yields $\partial_r \beta rB_\phi=0$ and can be integrated to give the
scaling $B_\phi \propto 1/r$ for relativistic flows. One can immediately see that the Poynting-flux 
term in equation (\ref{lum}) is approximately constant and no further acceleration of the
flow is possible within ideal MHD for a radial flow. This is a result of the fact that
the magnetic pressure and tension terms of the Lorentz force almost cancel each other (Begelman 
\& Li 1994).   

We argue, however, that when the toroidal component of the magnetic field becomes dynamically dominant 
the kink instability sets in. The instability drives its energy from $B_\phi^2$ on the instability
growth time scale. This effect can be crudely modeled by the addition of one sink term on the right 
hand side of the induction equation following Drenkhahn (2002), Drenkhahn \& Spruit (2002)
\be
\partial_r \beta rB_\phi=-\frac{rB_\phi}{ct_{\rm{k}}}.
\label{induction}
\ee  
The kink instability time scale is given by the expressions (2) or (1) depending on whether
the poloidal component of the magnetic field is assumed to have a stabilizing effect. 
When the instability sets in, $B_\phi$ drops faster than $1/r$ and acceleration of the flow is
possible at the expense of its magnetic energy.  

\subsection{Radiative losses}

The dynamical equations  (\ref{energyc}) and (\ref{momentumc})
are derived under the assumption that no energy or momentum escape from the
outflow. This is accurate when the instability releases energy in
the optically thick region of the flow. On the the other hand,  
in the optically thin regime energy and momentum may be transfered into the 
radiative form that escapes and does not interact with matter.  Let $\Lambda$ be
the emissivity of the medium in the comoving frame, that is, the energy
that is radiated away per unit time and per unit volume.  If the
emission is isotropic in the comoving frame the energy and momentum
Eqs.~(\ref{energyc}), (\ref{momentumc}) including the radiative
loss terms are (K\"onigl \& Granot 2002)
\begin{equation}
  \label{energyc2}
  \partial_r r^2 \left(
    w \gamma u + \frac{\beta B_\phi^2}{4\pi}
  \right) 
  = - r^2 \Gamma\frac{\Lambda}{c}
  \ ,
\end{equation}
\begin{equation}
  \label{momentumc2}
  \partial_r r^2 \left(
    w u^2 + p
    + \left(1+\beta^2\right) \frac{B_\phi^2}{8\pi}
  \right) 
  = 2rp - r^2 u\frac{\Lambda}{c}.
\end{equation}

The importance of the cooling term depends on the cooling time
scale.  If it is short compared to the expansion time scale, the matter stays 
cold during the dissipation process.  In this limit, all the
dissipated energy is locally radiated away. The dissipative processes
that appear in the non-linear stage of the instability are poorly understood.
It could be the case that the released energy leads to fast moving particles
(i.e. electrons and ions) and/or to Alfv\'en turbulence (Thompson 1994).   
Synchrotron emission is a plausible fast cooling process for the electrons.  It is particularly 
effective in our model, because the magnetic field strengths are high in a Poynting
flux dominated outflow. Ions, however, are, due to their higher masses, much less efficient
radiators.   
    
The form of this cooling term we assume here is
\begin{equation}
  \label{eq:Lambda}
  \Lambda
  = \kappa \frac{ecu}{r}
\end{equation}
where $k$ is an adjustable cooling length parameter.  The cooling
length is the distance by which the matter travels outward while the
internal energy $e$ is lost. When $\kappa\gg 1$ the cooling length is very short, only a small
fraction of the expansion length scale $r$ and thus qualifies for the
description of a fast cooling flow. This, in more physical terms, corresponds to the case
where most of the energy is dissipated to fast moving (and therefore fast cooling) electrons. On the other
hand, setting $\kappa\ll 1$, the cooling length
is much longer than the expansion length and most of the energy stays in the flow 
leading to more efficient adiabatic expansion. This is the case when the dissipated energy is mostly 
shared among the ions.  

\subsection{Initial conditions, model parameters}

The characteristics of the flow are determined when a number of quantities are specified at the 
fast magnetosonic point $r_0$ which is taken to be $\sim$ a few times the light cylinder radius
(Sakurai 1985; Begelman et al. 1994; Beskin et al. 1998),
or expressed in terms of the gravitational radius $r_g=GM/c^2$ of the central engine 
$r_0\sim 100r_g$. These quantities are the initial magnetization $\sigma_0$, the
luminosity $L$, the opening angle $\theta$, the ratio $B_{r,0}/B_{\phi,0}$ and the
cooling length scale $\kappa$.  
The quantities one has to solve for so as to determine the characteristics of the flow
are $\rho$, $e$, $u$ and $B_\phi$ as functions or radius $r$. This is done by integrating
numerically the mass, energy, momentum conservation equations and the modified induction equation.
The parameters of the model determine the initial values of $\rho$, $e$, $u$ and $B_\phi$ at
$r_0$.  

The initial four-velocity for our calculations is assumed to be
\be
u_0=\mu^{1/3}=\sqrt{\sigma_0},
\ee 
in accordance with previous studies (Michel 1967; Goldreich \& Julian 1970; Camenzind 1986;
Beskin et al. 1998) which show that at the fast point the ratio of Poynting to kinetic flux is
$\mu^{2/3}$. The flow is assumed to be cold at $r_0$, i.e. $e=0$ and using the previous
expression with Eqs.~(\ref{lum}), (\ref{sigma}) one finds for  $\rho_0$ and $B_{\phi, 0}$ 
\be
\rho_0=\frac{L}{r_0^2c^3\sqrt{\sigma_0(\sigma_0+1)^3}}
\ee
and
\be
B_{\phi,0}=\frac{4\pi}{r_0}\Big(\frac{\sigma_0}{\sigma_0+1}\Big)^{1/4}\sqrt{\frac{L}{c}}.
\ee

 The role of the different model parameters becomes
clear in the next section where the model is applied to the case of AGN jets and GRBs.
Out of the free parameters of the model, $\sigma_0$ and $\theta$ are of special 
importance. The magnetization
$\sigma_0$ determines the ``magnetic dominance'' of the flow, i.e., the speed of the flow at the
fast magnetosonic point and at a large distance from the central engine.
On the other hand, the opening angle $\theta$ is directly related to the growth rate of the 
instability [see Eqs. (1), (2)]. The 
instability has enough time to grow if it is faster than the expansion
time $r/c$. The ratio of the two time scales at the base of the flow is (using prescription (1)
for the time scale of the kink instability)
\be
t_k/t_{exp}=\frac{\theta \gamma_0c}{v_{A,\phi}}\simeq \frac{\theta \sqrt{\sigma_0}c}{v_{A,\phi}}.
\label{ratio}
\ee      
If $t_k/t_{exp}\gg 1$, the kink instability does not
have enough time to grow and the evolution is close to that predicted by ideal MHD (where
not much acceleration takes place). On the other hand, if  $t_k/t_{exp}\ll 1$, the instability
grows for many e-foldings and turns almost all the magnetic energy in the flow into
radiation and kinetic flux. Keeping the opening angle fixed, this happens much more
efficiently in lower $\sigma_0$ flows (provided that $\sigma_0\simmore 1$ so that
the Alfv\'en speed is a significant fraction of the speed of light). We return to this point in the
next sections.

\section{Applications}

Although at first sight different, jets in AGNs (and microquasars) and GRBs probably have central
engines of similar characteristics. AGN jets are launched in the inner regions
of magnetized accretion disks (Blandford \& Payne 1982), or drive their power by magnetic fields that are  
threading the ergosphere of a rotating black hole (Blandford \& Znajek 1977). In the case of GRBs,
the same central engine may be at work, or the energy is tapped by a millisecond magnetar
(Usov 1992; Klu\'zniak \& Ruderman 1998; Spruit 1999). In all of these situations, 
strong magnetic fields play an important role
and most of the energy is released in the form of Poynting flux. \footnote{ An exception is the
possibility of creation of a fireball by neutrino-antineutrino annihilation at the 
poles of a hyperaccreting compact object (Jaroszy\'nski 1993; Mochkovitch et al. 1993), an
idea applied to long bursts within the collapsar scenario (Woosley 1993; MacFadyen \& Woosley
1999) and short bursts within the binary merger scenario (Blinnikov et al. 1984; Eichler et al.
1989; Janka et al. 1999; Aloy et al. 2005)} 

All the above scenarios may give rise to magnetized outflows, whose evolution
depends, to a large extent, on the dominance of the magnetic energy or on the ratio 
of the Poynting-flux to matter energy flux at the base of the flow.
By varying this ratio, one can apply the model to jets in both the cases of GRBs and AGNs.

\subsection{AGN jets}

Relativistic jets are commonly observed in AGNs to have bulk Lorentz factors in the range 
$\gamma\sim 10-20$. Such terminal Lorentz factors can be achieved for the ratio
$\sigma_0$ of Poynting to matter energy flux of the order of several at the fast magnetosonic 
Point $r_0$. The location of the fast point is most likely at a few light cylinder radii
(e.g. Sakurai 1985; Camenzind 1986) and is taken to be $100r_g$.
 
Actually, $\sigma_0$ is a very important parameter of the flow. Its effect on the acceleration
of the flow is clearly seen in Fig. 1, where the bulk Lorentz factor is plotted as a function
radius $r$ for different $\sigma_0$. The rest of the parameters have the values $\theta=10^o$,
$B_{r,0}/B_{\phi,0}=0.5$, while the energy released by the instability is assumed to be locally radiated
away (this is done by taking the ``cooling length'' parameter $\kappa\gg 1$). The results do not
depend on the luminosity $L$ of the flow in the case of AGN jets, while $r_0$ sets the scale
of the problem (since it is the only length scale) which means that the results can be trivially
rescaled in the case of a different choice of $r_0$.
 
\begin{figure}
\resizebox{\hsize}{!}{\includegraphics[angle=270]{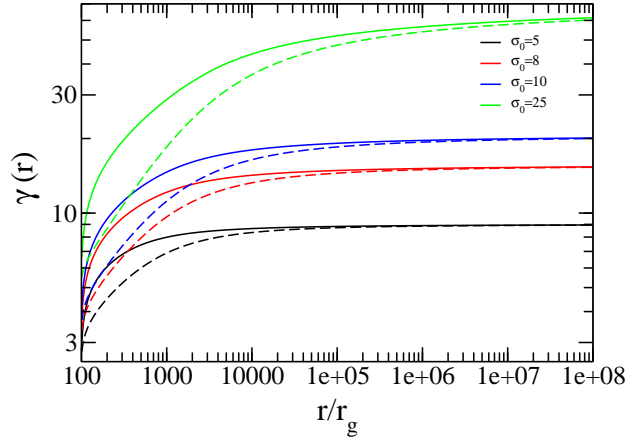}}
\caption[]
{The bulk Lorentz factor of the flow as a function of radius for different
values of $\sigma_0$. The black, red, blue and green curves correspond to $\sigma=5,$ 8, 10
and 25 respectively. The solid curves correspond to the case where Eq.~(1) is used for the
instability growth time scale (fast kink) and the dashed to the one where Eq.~(2) is used
(slow kink case).  
\label{fig1}
}
\end{figure}

The solid lines in Fig. 1 correspond to the case where Eq.~(1) is used for the timescale of the kink 
instability (i.e., the fast kink case) and the dashed lines to the case where 
the instability is slowed down by the poloidal component of the magnetic field and the time scale is 
given by Eq. (2) (i.e., the slow kink case). 
From Fig.~\ref{fig1}, one can see that the instability acts quickly and accelerates the flow
within 1-2 orders of magnitude in distance from the location of the fast magnetosonic point. 
The acceleration is faster in the ``fast kink'' case and much more gradual in the ``slow kink'' one. 
This is due to the fact that close to the base of the flow the ratio 
$B_{r}^{co}/B_{\phi}^{co}=\gamma B_{r}/B_{\phi}\sim 1$ and the instability is slowed down [see Eq. (2)].
Further out, however, the toroidal component of the field also dominates in the frame comoving with the
flow and the instability proceeds faster. At larger distances, practically all the magnetic energy
has been dissipated and the terminal Lorentz factors are very similar in the slow and fast 
kink cases.

%-----------------------------------------------------------------  
%----------------------------------------------------------------- 

The acceleration of the flow and the terminal Lorentz factor depend also on what
fraction of the instability-released energy is radiated away. If the dissipative processes
that appear in the non-linear regime of the evolution of the instability lead to fast
moving electrons, then it is easy to check that they will radiate away most of this energy
through synchrotron (and/or inverse Compton) radiation on a time scale much shorter than the
expansion timescale. If, on the other hand, most of the energy is dissipated to the ions,
then most of it stays in the system as internal energy and accelerates the flow further.     
To keep this study fairly general, we have calculated the bulk Lorentz factor of the
flow in the two extreme cases. In the ``fast cooling'' case, all the released energy is
radiated away very efficiently, while in the ``slow cooling'' case, the energy is assumed
to stay in the flow (practically this means that we set the cooling length parameter $\kappa\ll 1$).

\begin{figure}
\resizebox{\hsize}{!}{\includegraphics[angle=270]{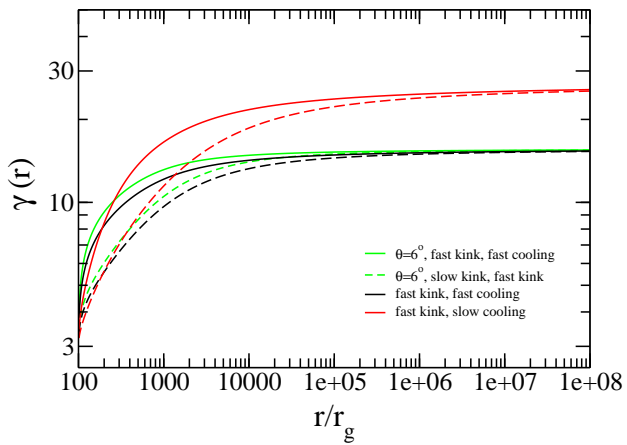}}
\caption[]
{The bulk Lorentz factor dependence on the cooling efficiency of the flow and jet opening angles. 
The solid curves correspond to the fast kink case while the dashed to the slow kink one.
The black, red and green  curves correspond to  fast cooling, slow cooling and jet opening angle
of $6^o$ respectively.
\label{fig2}
}
\end{figure}

In Fig.~\ref{fig2}, the bulk Lorentz factor of the flow is plotted for $\sigma_0=8$. The red curves
correspond to the ``fast cooling'' case and the black to the ``slow cooling'' one. The asymptotic bulk 
Lorentz factor differs substantially in these two cases, showing that a large fraction of the
energy of the flow can in principle be radiated away due to the instability-related dissipative 
processes. Furthermore, the acceleration of the flow depends on the jet opening angle and is
faster for narrower jets (see green curves in Fig.~\ref{fig2}). This is expected, since for a narrower 
opening angle, the Alfv\'en crossing time across the jet is shorter and so is the instability growth 
timescale.

%-----------------------------------------------------------------  
%----------------------------------------------------------------- 

Another quantity of special interest is the Poynting to matter energy flux ratio $\sigma$
as a function of radius. While the flow is initially moderately Poynting flux
dominated, the $\sigma$ drops rapidly as a function of distance and the flow is matter-dominated 
at distances $r\simmore 10^3r_g$ independently of the prescription of
the instability or the cooling timescales. Far enough from the fast magnetosonic point,
practically all the magnetic energy has been transfered to the matter, and the bulk
Lorentz factor saturates. 

\begin{figure}
\resizebox{\hsize}{!}{\includegraphics[angle=270]{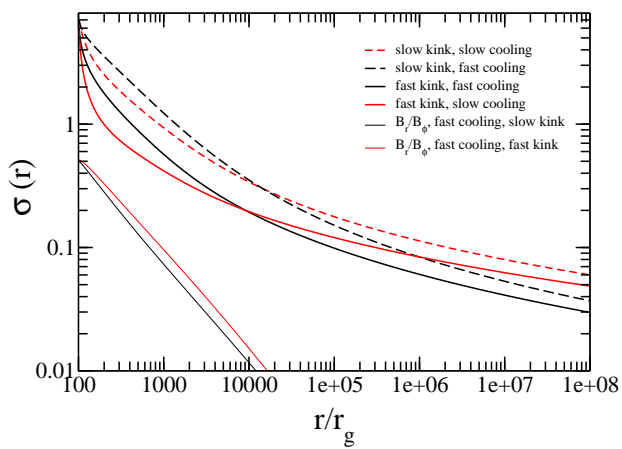}}
\caption[]
{The dependence of the magnetization parameter on the radius  
for the different prescriptions of radiative cooling and the instability
timescale. Notice that the flow becomes matter-dominated at distances greater than 
$\sim 10r_0$. The thick lines show the ratio of the radial to the toroidal
components of the magnetic field in the flow. The dominant component is clearly the
toroidal one. 
\label{fig3}
}
\end{figure}

The thick lines in Fig.~\ref{fig3} show the ratio of the radial to the
toroidal components of the magnetic field. This ratio drops rapidly as a function of 
distance showing that $B_\phi\gg B_r$. So, despite the fact that the
instability grows quickly from the toroidal component of the magnetic field,
this component still dominates over the radial one.  

%-----------------------------------------------------------------  
%-----------------------------------------------------------------     

Having solved for the dynamics of the flow predicted by our modeling of
the kink instability, we turn to the implications of these findings 
for observations of AGN jets. One highly debated issue is
whether the AGN jets are Poynting-flux dominated on pc and kpc scales
or not. The case-dependent arguments are reviewed in Sikora et al. (2005),
where it is shown that there is no strong observational reason to assume
Poynting-flux dominated jets on scales larger than a few pc and that the 
observed emission on these scales can be understood as energy
dissipated in shocks internally in the flow (Sikora et al. 1994; Spada et al.
2001) or due to interaction of the flow with the external medium.           

Our model predicts that most of the energy is in the form of kinetic
flux at distances say $\simmore 10^3-10^4 r_g\simeq 10^{17}-10^{18}m_9$ cm, where
$m_9$ is a black hole of $10^9$ solar masses. So, on pc scales the 
magnetic fields are dynamically insignificant, in agreement with observations.
Further information on the dynamics of AGN jets comes from the shortest variability
timescale in the optical and gamma-ray bands in blazars. This timescale can
be as short as a few days, indicating that most of the non-thermal radiation comes
from a compact region of size $R\simless 10^{17}$ cm (the so-called blazar zone).
On the other hand, polarimetry measurements of the variable optical, infrared and mm 
radiation are consistent with a toroidal magnetic field geometry on sub-pc scales
(e.g. Impey et al. 1991; Gabuzda \& Sitko 1994; Nartallo et al. 1998).

Since most of the magnetic energy is dissipated on these scales, it is quite 
probable that the observed radiation is the result of the instability-released energy,
provided that the dissipative processes lead to wide enough particle
distributions (see also Sikora et al. 2005). However, one cannot exclude the possibility 
that, within this model, the ``blazar zone'' emission is a result of internal shocks. 
On scales of $10^{17}$ cm, the magnetization parameter of the flow is of the order of unity 
and it is interesting to study the outcome of internal shocks of moderately magnetized plasma.  
The rich blazar phenomenology may indicate that both these mechanisms (i.e. magnetic 
dissipation and internal shocks) are at work.        

Further constraints on where the acceleration of the flow takes place come from the lack of 
bulk-flow Comptonization features in the soft X-rays.
This indicates that $\gamma\simless 10$ at $\sim 10^3 r_g$ (Begelman \& Sikora 1987;
Moderski et al. 2003). 
This shows that the acceleration process is still going on at these distances. In view of 
our results, this could in principle rule out the ``fast kink'' case since the acceleration 
appears to be too fast and $\gamma\sim 10$ already at $\sim 300 r_g$ or so. At this point, however, 
the uncertainties in the model are too high to make a strong statement on this issue. 
If, for example, the fast point is located at a factor of, say, $\sim 3$ larger distance, 
our results  are compatible with the lack of soft X-ray features. Numerical simulations of the instability
are needed so that these issues can be settled (see also discussion in Sect. 5).

\subsection{Gamma-ray bursts}

The analysis we follow so as to apply the model to GRBs is very similar
to that described in the previous sections. The only new ingredient 
that has to be added is related to the very high luminosities that
characterize the GRB jets. As a result, the inner part of the flow
is optically thick to electron scattering and matter and radiation are
closely coupled. At the photospheric radius the optical depth 
drops to unity and further out the flow is optically thin. 
So, the high luminosity introduces a new length scale to the 
problem  that has to be treated in a special way described in the next section. 

\subsubsection{Below and above the photosphere}

At the photosphere, the equation of state changes from one dominated
by radiation to one dominated by the gas pressure.  To connect the
two, the radiation emitted at the photosphere has to be taken into
account.  The amount of energy involved can be substantial, and
appears as an (approximate) black body component in the GRB spectrum.
It depends on the temperature of the photosphere.

The temperature at the photosphere is $kT\ll m_e c^2$ for all 
parameter values used so that pairs can be neglected.  The photosphere is
then simply defined as (e.g. Abramowicz et al. 1991)
\be
\int_{r_{\rm{ph}}}^\infty (1-\beta)\gamma \kappa_{es}\rho \rm{d}r\equiv 1,
\label{rph}
\ee
where $\kappa_{es}$ is the electron scattering opacity. 
The dynamics of the flow depend on the location of the photosphere
since above the photosphere all the dissipated energy can in principle
be radiated away while this is not possible at large optical depths.
To solve Eq.~(\ref{rph}) for $r_{\rm{ph}}$, we have followed an 
iterative method. First, we guess a value for $r_{\rm{ph}}$ and integrate
the dynamical equations assuming no radiative losses below the photosphere
and fast cooling above it. Then we calculate the optical depth $\tau$ from $r_{ph}$
to $\infty$ and, if $\tau$ differs from unity by more than a threshold value
($\sim 0.01$ in these calculations), a new guess for $r_{\rm{ph}}$ is made.
The procedure continues until the definition (\ref{rph}) is satisfied.      

At the photosphere one has to subtract the energy and momentum  carried away 
by the decoupled radiation.  To calculate these
quantities one needs the temperature at the photosphere.
The dimensionless temperature $\theta=kT/(m_e c^2)$ in the optically
thick region is given by the solution of
\begin{equation}
  \label{eq:theta}
  e
  = 3 \frac{m_\mathrm{e}}{m_\mathrm{p}} \rho c^2 \theta
  + \frac{8\pi^5}{15} \frac{m_\mathrm{e}c^2}{\lambda_e^3} \theta^4
\end{equation}
where $\lambda_e$ is the electron Compton wave length. The terms in 
(23) correspond to the matter and radiation energy density. From this solution,
we have also found that the internal energy is always dominated by radiation 
(for parameters relevant for GRBs), so we take $\gamma_a=4/3$ below the photosphere.

At the photosphere we calculate the temperature $\theta_\mathrm{ph}$
and subtract the radiation energy density of a black body
\begin{equation}
  \label{eq:ebb}
  e_\mathrm{bb}
  = \frac{8\pi^5}{15} \frac{m_\mathrm{e}c^2}{\lambda_e^3}
  \theta_\mathrm{ph}^4
\end{equation}
from the total energy density: $e \equiv e - e_\mathrm{bb}$.  The
integration proceeds with an adiabatic index of $\gamma_a = 5/3$.  The
temperature $\theta_\mathrm{ph}$ is the temperature of the emitted
black-body radiation which has a luminosity per sterad of
\begin{equation}
  \label{eq:Lbb}
  L_\mathrm{ph}
  =       r_\mathrm{ph}^2 \frac{4}{3} e_\mathrm{bb}
      u_\mathrm{ph} \gamma_\mathrm{ph} c  \quad \mbox{for~} r\ge r_\mathrm{ph}.
\end{equation}

The integration continues until large distances from the source (taken as $10^{16}$ cm, where
the afterglow phase starts).  There, the
radiative luminosity is determined by
\begin{equation}
  \label{eq:Lnt}
  L_\mathrm{rad} 
  = L - L_\mathrm{pf} - L_\mathrm{mat} 
  \ .
\end{equation}
This means that the radiative luminosity is the sum of the photospheric luminosity
plus the component coming from  the instability-released energy above the photosphere.
The role of the photospheric component and its connection to the observed spectral
peaks of the GRB prompt emission in internal shock and slow dissipation models
(like this one) has been studied in a number of recent papers (Ryde 2005;
Rees \& M\'esz\'aros 2005; Pe'er et al. 2005). 

\subsubsection{Results}

Following the procedure described in the previous section, we have 
calculated the bulk Lorentz factor of the flow for different
values of $\sigma_0$ at the base of the flow and for the two
prescriptions for the timescale of the kink instability [Eqs.~(1), (2)].
The fast magnetosonic point is set to $R_0=10^8$ cm, the jet opening angle
to $\theta=10^o$, the initial ratio $B_{r,0}/B_{\phi,0}=0.5$ and the luminosity
of the flow $L=10^{51}\quad \rm{erg}/\rm{sec\cdot sterad}$. 

\begin{figure}
\resizebox{\hsize}{!}{\includegraphics[angle=270]{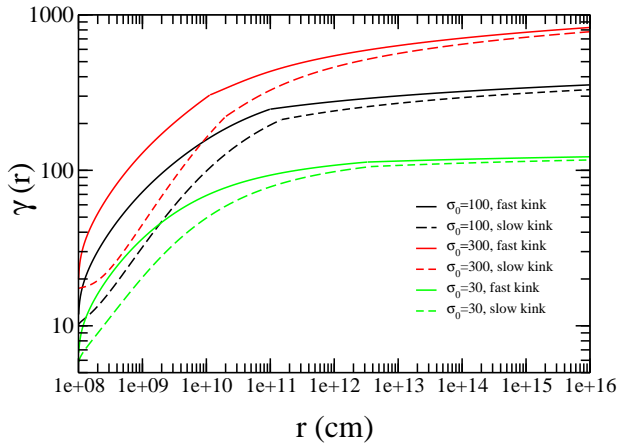}}
\caption[]
{The bulk Lorentz factor of the flow for different
$\sigma_0$ and for the fast and slow kink case. Notice that larger 
values for $\sigma_0$ result in faster outflows. 
\label{fig4}
}
\end{figure}

The results are given in Fig.~\ref{fig4}, where it is shown that the flow
reaches terminal Lorentz factors $\gamma\simmore 100$ for $\sigma_0 \simmore 30$.
 The solid curves correspond to the case where the timescale for the kink
instability is given by Eq.~(1) (fast kink case) and the dashed to the case 
where Eq.~(2) is used for the timescale of the growth of the instability 
(slow kink case). Notice that the initial acceleration of the flow differs 
in the two cases, being much faster in the fast kink case. This is expected 
since this case is characterized by rapid dissipation of magnetic
energy and acceleration from the base of the flow, while in the slow kink case
the non-negligible poloidal component of the magnetic field close to $r_0$ slows
 down the instability. The terminal Lorentz factors are, however, similar
in the two cases. Notice also that there is a discontinuity in the slope 
of the curves $\gamma(r)$ at the location of the photosphere which is a result
of our simplistic approach (for details see previous section).

%-----------------------------------------------------------------  
%-----------------------------------------------------------------     

\begin{figure}
\resizebox{\hsize}{!}{\includegraphics[angle=270]{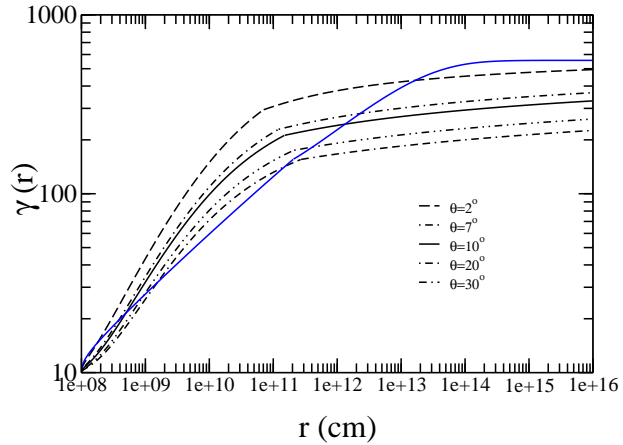}}
\caption[]
{The bulk Lorentz factor of the flow for different jet opening angles
$\theta$. For smaller opening angles, the terminal Lorentz factors of the flow become
larger because of more efficient dissipation of the magnetic energy. 
The blue curve corresponds to the non-axisymmetric case studied by Drenkhahn \& Spruit (2002).  
\label{fig5}
}
\end{figure}

\begin{figure}
\resizebox{\hsize}{!}{\includegraphics[angle=270]{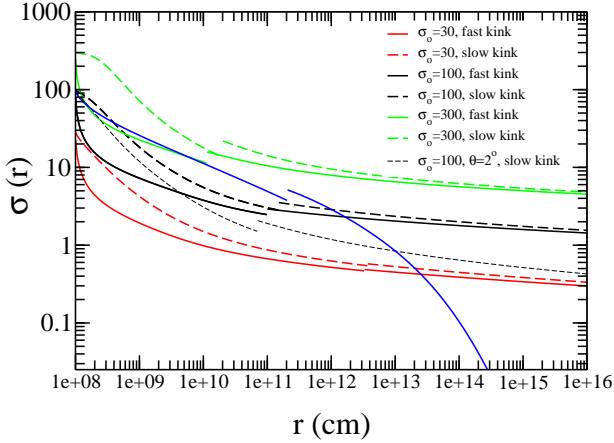}}
\caption[]
{The magnetization $\sigma(r)$ of the flow as a function of distance for
different $\sigma_0$ and jet opening angles. Keeping $\theta=10^o$, the jet is
still magnetically dominated at large distance from the source for $\sigma_0\simmore 100$.
Smaller jet opening angles lead to lower values of $\sigma_\infty$. 
The blue curve corresponds to the non-axisymmetric case studied by Drenkhahn \& Spruit (2002). 
The discontinuity at the location of the photospheric radius is a result of the subtraction of
the radiation energy density from the internal energy of the flow. 

\label{fig6}}
\end{figure}

A second key parameter of the model is the opening angle of the jet. 
For smaller opening angles, the instability timescale becomes
shorter and the flow is accelerated faster and to higher terminal
Lorentz factors as is shown in Fig.~\ref{fig5}. This implies that
for smaller opening angles, more magnetic energy is dissipated 
and the flow is less strongly magnetized at large distances. 
This is clearly shown in Fig.~\ref{fig6}, where the Poynting to
matter energy flux ratio is plotted as a function of radius $r$
(compare the thin curve with the thick black dashed curves).
Notice that the $\sigma$(r) curves are discontinuous at the location of
the photosphere. This is caused by our simplified treatment at the
location of the photosphere, where we subtract the energy density
of the radiation field (see previous section) and reduce the internal
enthalpy of the flow, increasing the ratio of Poynting to matter energy
flux. More detailed radiative transfer models of the transition from optically thick
to optically thin condition, predict a rather sharp transition which 
indicates that our simple approach does not introduce large
errors.  
    
In Figs.~5 and 6 we have also plotted (see blue curves) the bulk Lorentz factor 
and the magnetization $\sigma$ as functions of $r$ for the ``typical values'' of the
parameters of the model proposed by Drenkhahn \& Spruit (2002; the ``AC'' flow). In the context of that 
model the magnetic field lines change direction on small scales and magnetic reconnection
dissipates magnetic energy and accelerates the flow. Notice that the non-axisymmetric model predicts
more gradual acceleration and rather higher terminal Lorentz factors (for the same
initial magnetization of the flow) than the current model. Furthermore, it is characterized by
efficient dissipation of the Poynting flux, resulting in negligible magnetization sufficiently far
from the central engine (at least in the case where the non-decayable axisymmetric component
is negligible).

%-----------------------------------------------------------------  
%-----------------------------------------------------------------     

One important point deduced from Fig.~\ref{fig6} is that, for $\sigma_0\simmore 100$,
the flow remains Poynting-flux dominated even at large distances away from the 
source where deceleration of the outflow because of its interaction
with the interstellar medium or the stellar wind is expected, which means 
that the instability is not fast enough to convert most of the
magnetic energy into bulk motion of matter. Afterglow observations
can in principle probe to the magnetic content of the ejecta through early
observations of the reverse shock emission (Fan et al. 2002; Zhang et al. 2003;
Kumar \& Panaitescu 2003). Modeling of the
forward and reverse shock emission in cases where quick follow ups were possible
suggests the existence of frozen-in magnetic fields in the ejecta (Kumar \& Panaitescu 2003)
that are dynamically important, with $\sigma \simmore 0.1$ (Zhang \& Kobayashi 2005).  
Rapid follow-ups in the X-rays, UV and optical  are now possible thanks to Swift 
satellite and ground based telescopes and can test our
model which predicts a magnetization parameter of the order of unity for the
outflowing material in the afterglow region. The XRT instrument on board Swift
has already provided several early X-ray afterglows (e.g. Tagliaferri et al. 2005; Campana et al. 
2005; Burrows et al. 2005). Many of these observations
indicate a slow fading component at times $~10^2-10^4$ sec after the GRB trigger (Nowsek et al. 
2005; Zhang et al. 2005; Panaitescu et al. 2005) which may be expected by the deceleration
of ejecta with $\sigma \simmore 1$ (Zhang \& Kobayashi 2005; Zhang et al. 2005) 
in agreement with our model predictions.  
  
%-----------------------------------------------------------------  
%-----------------------------------------------------------------     

The ratio $\sigma$  is even higher for the range of distances $r\sim 10^{13}-10^{15}$ cm 
where internal shocks are expected to happen in the internal shock scenario for GRBs (Rees \&
M\'esz\'aros 1994; Piran 1999) and is expected the reduce their radiative efficiency. 
However, allowing for non-ideal MHD effects in the shocked region, Fan et al. (2004) show that 
the radiative efficiency of $\sigma\sim 1$ plasma may not be much lower than 
the $\sigma=0$ case.  On the other hand, since the efficiency of internal 
shocks to convert kinetic energy into gamma rays is already low (typically
of the order of a few percent; Panaitescu et al. 1999; Kumar 1999) and observations indicate
much higher radiative efficiency (Panaitescu \& Kumar 2002; Lloyd \& Zhang 2004), we investigate
the possibility that the energy released by the instability powers the prompt gamma-ray emission.  

\begin{figure}
\resizebox{\hsize}{!}{\includegraphics[angle=270]{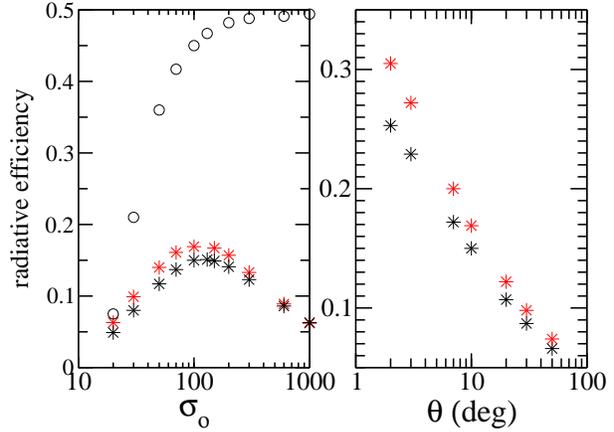}}
\caption[]
{The radiative efficiency of the flow, defined as the ratio of the radiated luminosity
over the luminosity of the flow for different $\sigma_0$ and $\theta$. The black and red stars
correspond to the fast kink and slow kink cases respectively. For opening angles of $\sim 6^o$
(in accordance with the values deduced by achromatic breaks of the afterglows) the efficiency
reaches values of $\sim 20$\%. The circles correspond to the non-axisymmetric case studied 
by Drenkhahn \& Spruit (2002).
\label{fig7}
}
\end{figure}

In Fig.~\ref{fig7}, we plot the
radiative efficiency -defined as the radiated luminosity $L_{\rm{rad}}$ divided by the flow
luminosity $L$- for different values of $\sigma_0$ and $\theta$. Fixing the angle 
$\theta$ to $10^o$, one can see that the radiative efficiency peaks at $\sim 16$\%
for $\sigma_0\sim 100$. For smaller values of $\sigma_0$, most of the magnetic energy is
dissipated below the photosphere and is lost to adiabatic expansion, resulting in lower
radiative efficiencies. For larger values of $\sigma_0$, the flow remains magnetically
dominated at all radii, keeping the radiative efficiency lower. Furthermore,
the ``slow kink'' case has rather higher efficiencies and this comes from the
fact that dissipation happens at larger radii and therefore in optically thin 
environments. So, the model can have large radiative efficiencies for 
$\sigma_0\sim 10-500$. Notice that one also needs $\sigma_0\simmore 30$ to 
overcome the ``compactness problem'' (e.g. Piran 1999).

Fixing $\sigma_0=100$, one can now calculate the radiative efficiency for different
opening angles of the flow. Smaller opening angles result in more magnetic energy
dissipated (by shortening the instability timescale) and therefore
to smaller values of Poynting to matter flux at large distances. This also means
that more energy is radiated away. Although very model dependent, the opening angles
of the GRB jets can be estimated by the achromatic breaks of the afterglow lightcurves
(Rhoads 1997, Sari et al. 1999). For $\theta\sim 6^o$ (a value typically inferred),
the efficiency is quite high and of the order of 20\%.   

In Fig.~7, the radiative efficiency of the
non-axisymmetric model (Drenkhahn \& Spruit 2002) is also shown for different values
of $\sigma_0$. The non-axisymmetric model can have a higher radiative efficiency
which is close to $\sim50\%$ for $\sigma_0\simmore 100$ (See also Giannios \& Spruit 2005
for more detailed study on the spectra expected from this model). 
 
%-----------------------------------------------------------------  
%-----------------------------------------------------------------     

\section{Discussion}

This work suggests that the kink instability plays a significant role in the dynamics
of magnetized outflows. The instability sets in once the toroidal component of the magnetic
field becomes dominant and drives its energy by $B_\phi$ on a short time scale. The energy dissipated
by the instability accelerates the flow and turns it into kinetic flux dominated flow
for AGN jets at distance $\simmore 1000r_g$ and to moderately magnetized flow for GRB jets in the
the afterglow region. If the dissipated magnetic energy is transferred to fast moving
electrons with wide enough energy distribution, then it can power the blazar zone emission
and the prompt GRB emission with high radiative efficiency. 

These results have been compared with those that are predicted by other
dissipative models (Coroniti 1990;  Spruit et al. 2001; Lyubarsky \& Kirk 2001; 
Drenkhahn 2002; Drenkhahn \& Spruit 2002). According to these models, if the magnetic field 
lines change direction on small 
scales, magnetic energy can be dissipated through reconnection processes. Drenkhahn (2002) and
Drenkhahn \& Spruit (2002) applied this idea to GRB outflows and showed that efficient acceleration
and radiation (as high as 50\%) is possible. In the context of this model, most of the magnetic energy
is dissipated, resulting in kinetic flux dominated flows at large distances where the flow starts
to be decelerated by the external medium. On the other hand, our model predicts moderately 
magnetized ejecta at this region. Since the initial phase of the afterglow emission depends
on the magnetic content of the ejecta (e.g. Lyutikov 2005), these models make different
predictions about this phase and can be tested against observations.

This study assumes a radial flow and although this allowed us to minimize the number
of free parameters and clarify the role of each of them, it nevertheless leaves a number 
of issues unsettled. Two important issues are these of jet collimation and of the non-linear
evolution of the kink instability. We discuss these issues in the next subsections.

\subsection{Collimation}

The collimation of MHD outflows is usually believed to take place in the trans-Alfv\'enic
region because of the ``hoop stress'' exerted by the toroidal component of the magnetic
field. One issue that arises is whether the same mechanism is at work in the case where
the kink instability sets in and reduces the strength of $B_\phi$.
 Our one dimensional approach cannot settle this question; 2-D calculations would be 
needed if the instability is parametrized as in the present models. Time dependent, 3-D
simulations will be needed if the effects of the instability are to be included realistically,
since the relevant ones are nonaxisymmetric. 
Collimation of the flow can be achieved by its interaction with the environment. 
This may be the collapsing star in the context of gamma-ray bursts (Woosley 1993) or a large scale
poloidal field in the case of AGN jets (Spruit et al. 1997). Another interesting possibility
is that small scale toroidal fields (probably a result of the development of the instability)
can lead to flow collimation under certain conditions (Li 2002). 

\subsection{The non-linear evolution of the instability}
 
The linear evolution of the kink instability is rather well understood and has
been studied by linearizing the MHD equations by a number of authors (e.g. Begelman 1998; 
Appl et al. 2000), which shows that the instability grows on the Alfv\'en crossing time across
the jet. The non-linear evolution of the instability is an issue that 
cannot be solved with analytical tools and 3-dimensional RMHD simulations that
cover many decades of radii are needed to settle this issue. Preliminary numerical
investigations have been done (e.g. Lery et al. 2000; Ouyed et al. 2003; Nakamura \& Meier
2004) which indicate that the kink instability is an internal mode that does not disrupt the
jet. On the other hand, whether it is able to rearrange the magnetic field configuration
internally in the flow on the short timescale implied by linear stability analysis is still
not clear. 

Some intuition on this issue can be gained by this study. We have tried two different
prescriptions for the instability growth time scale, the second of which accounts for its
possible slowing down because of a strong poloidal ``backbone'' in the core on the jet
(Ouyed et al. 2003). A non-negligible poloidal component can slow down the
initial growth of the instability; eventually it grows in a conical jet. This occurs 
because as the jet expands, the $B_\phi$ and $B_p$ scale as $1/r$ and $1/r^2$ respectively so as to
satisfy the induction equation. This means that the toroidal component
dominates the poloidal at some point and the instability sets in. A study that assumes a cylindrical jet,
 on the other hand, will not deal with the $B_\phi \gg B_p$ situation. Since the observed jets
do expand laterally (despite their strong collimation)  by many orders in radius from their
launching region to their termination shock, we believe that it is important for numerical
investigations of the role of kink instability to allow for jet expansion to reveal the 
characteristics of the non-linear development of the instability.

\subsection {More realistic models}

The limitations of the calculations presented here are obvious from the
parameterizations used. One may wonder to what extent these can be overcome in
numerical simulations. Since the most relevant instabilities are
nonaxisymmetric, such simulations have to be 3-dimensional. The computational expense of 3D
MHD simulations puts strong limitations on the kind of calculations that can be
done, and the realism of the conclusions that can be drawn from them. An
astrophysical jet operates over many decades in length scale, with different physics
dominating at different distances from the source.

For reasons of computational feasibility, the 3D simulations that have been
done so far use only a small range in distance, or a cylindrical geometry
(e.g. Nakamura et al. 2001; Ouyed et al. 2003; Nakamura \& Meier  2004). 
In the first case, the range of distance 
is too narrow to follow the consequences of 3-D instabilities effectively. 
In the second case, the effect of instability is limited by the boundaries. 
It is well known that kink instability can saturate into a finite amplitude, 
helical equilibrium when confined in a cylinder (in the astrophysical context 
see e.g. K\"onigl and Choudhuri 1986; Lyubarskii 1999).

But a computational cylinder taylored to the size of the source covers a
negligible range in length scales perpendicular to the axis, compared with an actual
jet. If, instead, the simulations were done in a spherical or conical geometry, the
continued expansion of the flow would stretch these helical configurations
perpendicular to the axis, immediately making them unstable again. This is the rationale for our
assumption that dissipation by instability will be a process that persists
for a large distance along the jet.

It may be possible to make numerical progress in, say, a conical
geometry, but limitations due to the finite range in length scales and time scales that
can be achieved will remain serious. For this reason, it is important to isolate
physical effects that can not (yet) be included realistically in simulations, and
explore them in more approximate models like the ones we have presented here.

\section{Conclusions}

The standard scenario for jet launching, acceleration  and collimation involves
large scale magnetic fields anchored to a rotating object (e.g. Blandford \& Payne 1982;
Sakurai 1985). The flow passes through 
three critical points, i.e. the slow, the Alfv\'en and the fast point. At the fast
point the ratio of Poynting to matter energy flux is much larger than unity
in the case of relativistic jets (Michel 1967; Camenzind 1986; Beskin et al. 1998)
while further acceleration of the flow appears hard to achieve within ideal MHD
except if the flow is decollimated (Li et al. 1992; Begelman \& Li 1994).
 
In this work, we study how this picture is modified when one takes into account the 
fastest growing current driven instability, i.e. the $m=1$ mode kink instability.
We have modeled the instability by modifying the induction equation to account for non-ideal
MHD processes and solving the relativistic MHD equations in the case of a radial flow. 
The instability is driven by $B_\phi$, dissipates Poynting flux and has been shown to be an efficient 
mechanism to accelerate the flow.    

The key parameter of the model is the ratio $\sigma_0$ of the Poynting to matter energy flux
at the base of the flow. A large part of the AGN jet phenomenology can be understood in the
context of this model for $\sigma_0\sim$ several. On sub-pc scales the flow is
Poynting-flux dominated with $B_\phi\gg B_r$. The flow is shown to be accelerated fast and to
become matter dominated already at $\sim$pc scales, while it reaches terminal bulk factors
of a few tens. The emission at the blazar zone can be a result of either internal shocks
that take place in an unsteady flow, where fast shells catch up with slower ones, converting
a small fraction of the bulk kinetic energy of the flow into radiation (Rees \& M\'esz\'aros 1994;
Spada et al. 2001), or direct manifestation of the energy released by the instability. 

Within the same model, we propose that GRBs are a result of more Poynting 
flux dominated outflows with $\sigma_0\sim$100. For these values of $\sigma_0$ the
flow reaches terminal bulk Lorentz factors of the order of a few to several hundreds, while
it remains moderately magnetized (i.e. $\sigma_\infty\sim 1$) at the afterglow region region.
Although there is evidence for magnetized ejecta from afterglow modeling (e.g. Kumar \&
Panaitescu 2003; Zhang \& Kobayashi 2005), more results are anticipated from early afterglow 
follow-ups that can test the model.

In the internal shock scenario for the prompt GRB emission, the shells collide at 
typical distances of $10^{13}-10^{15}$ cm, where the flow is moderately Poynting-flux
dominated.  On the other hand, internal shock and Poynting-flux models exclude each other somewhat. If a
strong magnetic field is added to an internally-shocked outflow, the radiative efficiency
is further reduced with respect to that expected from the collision of unmagnetized shells
(e.g. Fan et al. 2004). At the same time, dissipation in a predominantly magnetic outflow by instability
(DC model) or internal reconnection (AC model) can produce radiation naturally at
very high efficiency (up to 50\%).

\begin{acknowledgements}
 Giannios acknowledges support from the EU FP5 Research Training Network ``Gamma Ray Bursts:
An Enigma and a Tool.'' 
\end{acknowledgements}

\end{document}